\documentclass[runningheads]{llncs}
\usepackage[T1]{fontenc}
\usepackage{graphicx}
\usepackage{longtable}   
\usepackage{booktabs}    
\usepackage{caption}     
\usepackage{array}       
\usepackage{tcolorbox}
\usepackage{subcaption} 
\usepackage{url}
\usepackage{multirow}    
\usepackage{amsmath}
\usepackage{float}  
\usepackage{enumitem}
\usepackage{cleveref}

\begin{document}
\title{Large Language Models for Fuzz Testing in Microservices: A Systematic Literature Review}
\titlerunning{LLMs for Fuzz Testing in Microservices: An SLR}
%
%

\authorrunning{Song et al.}

\author{Ying Song \inst{1}\orcidID{0000-0001-9791-1879} 
\and
Ke Ping \inst{1}\orcidID{0009-0009-5433-4771} 
\and
Yuqing Wang\inst{1}\orcidID{0000-0003-0175-005X} \and
Xiaozhou Li \inst{2}\orcidID{0000-0002-3767-2527}\thanks{Corresponding author.}}


\institute{
University of Helsinki, Helsinki, Finland \\
\email{firstname.lastname@helsinki.fi}
\and
Free University of Bozen-Bolzano, Bolzano, Italy \\
\email{xiaozhou.li@unibz.it}
}

\maketitle           
\begin{abstract}

Microservice systems (MSS) increasingly rely on heterogeneous APIs whose combinatorial input space and stateful dependencies challenge traditional fuzz testing. Meanwhile, Large Language Models (LLMs) have recently been introduced to enhance fuzzing with semantic reasoning over specifications, inputs, and runtime feedback. This paper presents a systematic literature review (SLR) of LLM-assisted fuzz testing for microservices to synthesise how LLMs are applied, evaluated, and what challenges remain. Following established SLR guidelines, we analyze 20 primary studies published between 2024 and 2026. Results show LLMs are mainly used as semantic input generators in black-box fuzzing, with a growing shift towards agent-based and retrieval-augmented architectures, improving valid input generation and modestly increasing coverage and vulnerability detection. However, evaluation remains heterogeneous, with limited benchmark standardization, scarce cost reporting, and a bias toward single-service experiments, highlighting a gap with real-world multi-service systems. 
This review provides a taxonomy of LLM roles and integrations, a consolidated view of evaluation practices, and a mapping of open challenges to research directions, supporting the design and deployment of LLM-driven fuzzing in microservices.

\end{abstract}
\section{Introduction}
\label{sec:intro}

Microservice architectures are widely adopted in modern software systems. They decompose applications into loosely coupled services that interact through APIs \cite{dragoni2017microservices}. While this improves scalability and maintainability, it also introduces significant testing challenges. In particular, the large parameter spaces of APIs, inter-parameter dependencies, and multi-step workflows across services make systematic testing in microservice systems (MSS) challenging \cite{wang2021promises,miao2025systematic}.

Fuzz testing has been widely used to address this problem, as it can automatically generate and execute diverse inputs to explore large numbers of test scenarios and detect failures such as crashes and server errors \cite{zhang2023open}. However, existing API fuzzers still have limitations. They often generate invalid inputs, struggle to satisfy complex constraints, and fail to reach deeper program states \cite{golmohammadi2023testing,manes2019art}. They also lack the ability to use API documentation, interpret error responses, or handle dependencies between inputs and business logic \cite{golmohammadi2023testing,manes2019art}. These limitations become more serious in MSS, where interactions span multiple services, depend on context, and are often not clearly specified \cite{S15_LogCoverage2026,S9_Wang2025bytedance}.


Recent advances in large language models (LLMs) provide new capabilities for understanding and generating information from diverse sources, such as API specifications, source code, logs, error messages, and request–response traces \cite{deng2023large}. These capabilities have been increasingly applied to fuzz testing, where LLMs are used to generate inputs, guide mutations, analyze responses, and coordinate testing workflows \cite{S9_Wang2025bytedance}. Although early results are promising, existing work is scattered across different domains, techniques, and evaluation settings, and has not been systematically analyzed. Despite the growing body of work, there is no consolidated understanding of how LLMs are applied to fuzz testing in MSS, or how effective these approaches are in practice. Existing secondary studies either focus on LLM-based fuzzing in other domains or on MSS testing without LLMs, leaving a clear gap at their intersection.

To address this gap, this paper presents a systematic literature review (SLR) of LLM-based fuzz testing in MSS. Following established guidelines for evidence-based software engineering
\cite{kitchenham2009systematic}, we employ a protocol-driven methodology that includes structured search, dual-reviewer screening, quality assessment, and data synthesis.
The SLR aims to systematically synthesise existing studies on the integration of LLMs into fuzzing for MSS, analyze how these approaches are evaluated and their reported effectiveness, and identify key limitations and research gaps.


The contribution of this study is threefold: (1) provides a structured taxonomy of LLM roles, fuzzing paradigms, and integration patterns in MSS testing; (2) consolidates evaluation practices across the literature, highlighting common metrics, baselines, and experimental limitations; and (3) identifies key research gaps, e.g., the lack of realistic multi-service benchmarks and cost-aware evaluation, and maps them to concrete future directions.


The remainder of the paper is organized as follows. Section~\ref{sec:related} positions our work against existing surveys. Section~\ref{sec:method} presents the methodology. Section~\ref{sec:results} 
reports the results. 
Section~\ref{sec:disc} discusses implications and open challenges. Section~\ref{sec:threats} outlines threats to validity. Section~\ref{sec:conclusion} concludes the paper.

\section{Related Work}
\label{sec:related}

Our work relates to two lines of secondary literature: SLRs on LLM-based fuzzing and SLRs on microservice testing. We discuss each and then identify the gap addressed by our review. 

\textbf{SLRs on LLM-based fuzzing.} 
Recent SLRs focus on LLM-based fuzzing tend to organise the literature by application domain rather than by fuzzing technique. For example, Huang et al.\cite{Huang2025} 
provide the most comprehensive review to date, surveying LLM-fuzzer 
integrations across binary, compiler, protocol, and DL-library 
targets, and characterising the roles LLMs play as seed generators, 
mutators, and harness synthesisers. Xu et al.\cite{xu2025largelanguagemodelscyber} review LLMs 
for cybersecurity tasks including web fuzzing, intrusion detection, 
and penetration testing. He et al.\cite{he2024large} focus on 
LLM-driven fuzzing of smart contracts, where the LLM guides exploration 
toward vulnerable code regions. Kaniewski et al.\cite{kaniewski2025systematic} 
review LLM-based software vulnerability detection more broadly, with 
fuzzing as one of several detection strategies.

\textbf{SLRs on microservice testing.} While MSS testing has been widely studied in SLRs, fuzzing and LLM-based techniques are not covered in these studies. For instance, Ghani et al.\cite{ghani2019microservice} and Waseem et 
al.\cite{Waseem} catalogue testing techniques, levels, and 
challenges in the microservice literature up to 2020, identifying 
integration and end-to-end testing as the dominant concerns and 
flagging the lack of standardised tooling.  Ponce et al.\cite{PONCE2025107870} update 
this picture with a more recent SLR of 74 primary studies organised 
along the SWEBOK testing taxonomy, reporting system testing as the 
most investigated level and highlighting the growing role of 
AI-assisted approaches without analysing them in depth. 


\textbf{Positioning.} The two lines of secondary literature above do not cover the intersection of LLM-based fuzzing 
and MSS.  LLM-based fuzzing reviews focus on other software domains, while MSS testing reviews do not focus on fuzzing or 
LLMs. Our SLR addresses this intersection: fuzz testing as a technique, LLMs as the AI component, and MSS as the system under 
test. 

\section{Methodology}
\label{sec:method}

\begin{figure}[!ht]
    \centering
    \includegraphics[width=\linewidth]{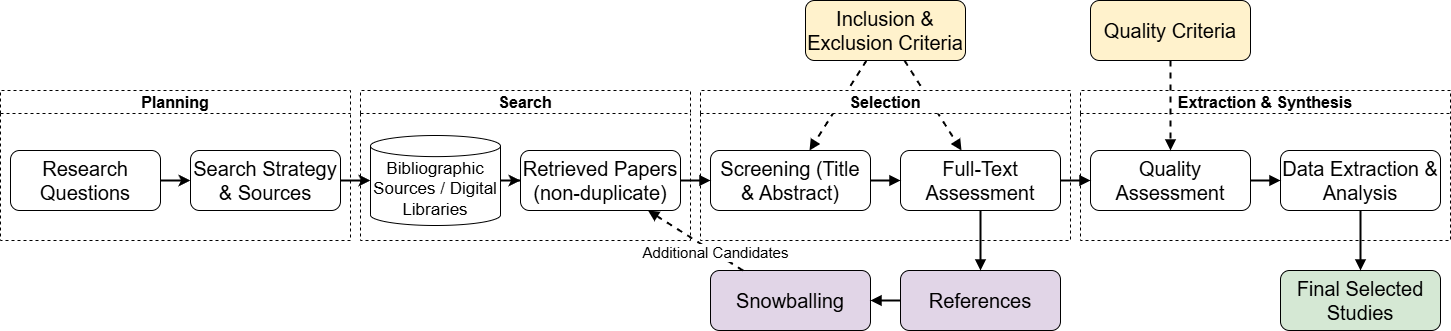}
    \caption{Systematic literature review (SLR) search and selection process}
    \label{fig:slrprocess}
\end{figure}

We follow the guidelines for systematic reviews in software engineering proposed by Kitchenham et al. \cite{kitchenham2009systematic} and complement them with Wohlin's snowballing procedure \cite{wohlin2014guidelines}. The goal is to identify, evaluate, and synthesise existing research on the use of Large Language Models (LLMs) for fuzz testing in microservice-based systems. The process of this study is depicted in Fig. \ref{fig:slrprocess}.

Consistent with established SLR practice, the review process comprises five stages: (1) Research question definition, (2) Search strategy and source selection, (3) Inclusion and exclusion screening, (4) Quality assessment, and (5) Data extraction \& synthesis. 

\subsection{Research Question Definition}

The research questions were defined to systematically characterise how LLMs are applied in fuzz testing for microservices and to assess the maturity of the field. In line with the guidance by Kitchenham et al., RQs are explicitly formulated to guide the search process, data extraction, and synthesis. The primary research questions are:

\begin{itemize}
    \item \textbf{RQ1.} How are Large Language Models applied to support fuzzing in microservice systems?
    \item \textbf{RQ2.} How are LLM-based approaches for microservice fuzzing evaluated, and what effectiveness do they achieve?
    \item \textbf{RQ3.} What are the open challenges and future research directions?
\end{itemize}

These RQs are motivated by the need to characterise the design space and assess the maturity of the field. Specifically, RQ1 classifies LLM roles and their integration with fuzzing paradigms such as black-box, gray-box, and white-box fuzzing. RQ2 examines evaluation practices, including target systems, metrics, baselines, and reported effectiveness. RQ3 identifies open challenges and maps them to future research directions. Together, these RQs support a three-part synthesis: (1) a descriptive mapping of techniques and design choices (RQ1), (2) an assessment of evaluation practices and reported effectiveness (RQ2), and (3) an analysis of open challenges and future research directions (RQ3). The structure is consistent with established SLR practices, where the RQs explicitly drive data extraction and analysis and align with the workflow shown in Fig. \ref{fig:slrprocess}.


\subsection{Search Strategy and Sources}

A comprehensive search strategy was designed to maximize the coverage of relevant studies while maintaining precision. Following best practices, the search combines automated database querying with a carefully constructed Boolean query. The final search string is structured along three dimensions:

\begin{enumerate}
    \item \textbf{LLM-related terms}: \begin{quote}\ttfamily\small\raggedright
    ("large language model*" OR LLM* OR "generative AI" OR "gen AI" OR "foundation model*" OR GPT* OR ChatGPT* OR Claude* OR Gemini* OR Llama* OR Copilot* OR Codex* OR DeepSeek* OR Grok* OR agent*)
    \end{quote}
    \item \textbf{Testing/fuzzing terms}: \begin{quote}\ttfamily\small\raggedright
    ("automated API testing" OR "REST API security testing" OR "REST API test*" OR "OpenAPI testing" OR "Swagger testing" OR fuzzing OR "fuzz testing" OR "fuzzer" OR "input generation" OR "test case generation" OR "automated testing")
    \end{quote}
    \item \textbf{System context terms}:  \begin{quote}\ttfamily\small\raggedright
    (microservice* OR "micro-service*" OR "micro service*" OR "service-oriented architecture" OR SOA OR "distributed system*" OR "web service*" OR API OR "RESTful" OR GraphQL)
    \end{quote}
\end{enumerate}

The search was conducted across major digital libraries commonly used in software engineering SLRs, including IEEE Xplore, ACM Digital Library, Scopus, Web of Science. These sources are consistent with Kitchenham et al.'s recommendation to search multiple repositories to reduce bias \cite{kitchenham2009systematic}. Furthermore, we also complemented the results with preprint articles from arXiv due to the fast-moving domain. In addition to automated database searches, we also employed backward and forward snowballing to improve coverage of relevant studies \cite{wohlin2014guidelines}. All studies identified through snowballing were subjected to the same inclusion/exclusion criteria and screening procedures.


\subsection{Inclusion and Exclusion Screening}

Explicit inclusion and exclusion criteria were defined to ensure consistency and reduce selection bias, as recommended in the SLR guidelines (Table \ref{tab:inex}). 

\begin{table}[!ht]
\centering
\caption{Inclusion and exclusion criteria.}
\label{tab:inex}
\renewcommand{\arraystretch}{1.15}
\begin{tabular}{p{2cm}p{10cm}}
\toprule
\textbf{Inc./Exc.} & \textbf{Criteria} \\
\midrule
\emph{Inclusion} & 
explicitly applies, proposes, or evaluates LLMs / generative AI / agentic AI for automated testing or fuzzing-related practices of API-based, service-oriented, distributed, or microservice systems \\
\midrule
\emph{Exclusion} & Irrelevant topic (no LLM \emph{or} no fuzzing/testing \emph{or} no MSS context). \\
 & Non-English source. \\
 & Inaccessible full text. \\
 & Studies that test LLMs/agentic-AI \emph{themselves} 
 \\
 & Published before 2023 (emergence of modern LLMs). \\
 & Other secondary studies 
 \\
\bottomrule
\end{tabular}
\end{table}

The inclusion and exclusion criteria were defined to ensure a precise alignment with the research scope while maintaining methodological rigor. In particular, the inclusion criterion directly addresses the RQs, while the exclusion criteria were designed to systematically remove sources that could introduce bias or irrelevance. This design follows established SLR practices that emphasize clearly defined, reproducible selection criteria to improve validity and consistency.

The screening process was conducted in two phases: (1) Title and abstract screening and (2) Full-text review. To improve reliability, each article was reviewed by two researchers with disagreements resolved by a third researcher. 

\subsection{Quality Assessment}

To evaluate the rigor and reliability of the included studies, a structured quality assessment was performed. The assessment is based on established criteria derived from the DARE framework used in previous SLRs \cite{kitchenham2009systematic}. The quality of each study was evaluated along four dimensions, including (1) clarity and appropriateness, (2) search strategy completeness, (3) validity assessment, and (4) reporting adequacy, with each criterion scored using a three-point scale. We set the selection threshold to be 6 out of 12. The assessment process was conducted by two reviewers with the quality score set by the average of two.



\subsection{Data Extraction and Synthesis}

A structured data extraction form was defined to systematically collect relevant information from each study. The extracted data aligns directly with the research questions and includes: Bibliographic information (title, venue, year), LLM characteristics (model type, usage), LLM role (e.g., input generator, mutation operator, oracle) Fuzzing type (black-box, grey-box, white-box), Target system (e.g., microservices, REST APIs), Evaluation setup (datasets, benchmarks, metrics), Availability of replication packages. Data extraction was performed by one researcher and validated by a second reviewer to ensure accuracy and consistency, following standard SLR practice \cite{kitchenham2009systematic}. For synthesis, we adopted a qualitative and descriptive approach, complemented by quantitative summaries where applicable, enabling both evidence aggregation and research gap identification.


\section{Results}
\label{sec:results}

We first describe the corpus (Section~\ref{sec:corpus}), then present the answers to RQ1 (Section~\ref{sec:rq1}), RQ2 (Section~\ref{sec:rq2}), and RQ3 (Section~\ref{sec:rq3}). Throughout, primary studies are cited with the labels \ref{SP1} - \ref{SP20} cross-referenced in Appendix.

\subsection{Corpus Overview}
\label{sec:corpus}

\begin{figure}[!ht]
    \centering
    \begin{subfigure}[t]{0.48\linewidth}
        \centering
        \includegraphics[width=\linewidth]{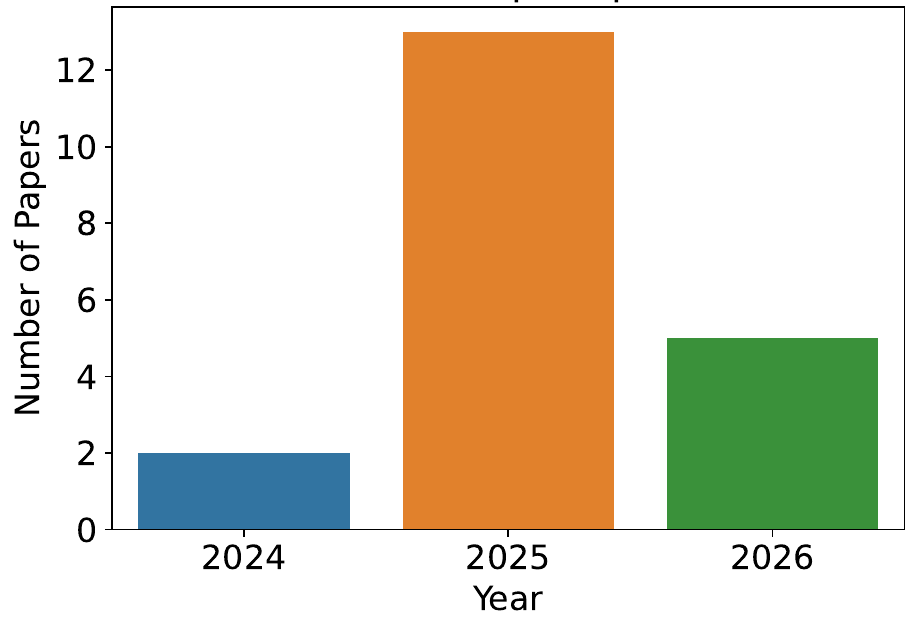}
        \caption{Number of papers per year}
        \label{fig:papers_per_year}
    \end{subfigure}
    \hfill
    \begin{subfigure}[t]{0.48\linewidth}
        \centering
        \includegraphics[width=\linewidth]{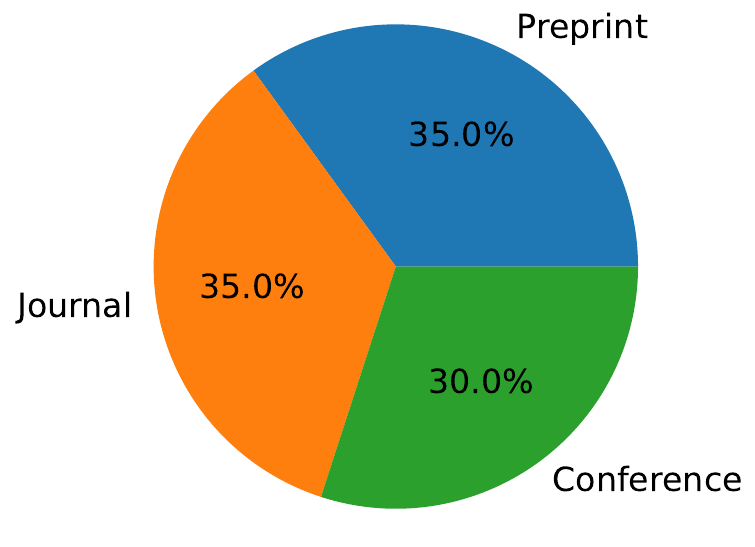}
        \caption{Distribution of document types}
        \label{fig:document_types}
    \end{subfigure}
    \caption{Overview of the selected studies: (a) yearly publication trend and (b) distribution of document types}
    \label{fig:slr_overview}
\end{figure}

The 20 included studies span 2024 (2 papers), 2025 (13 papers) and 2026 (5 papers), reflecting that LLM-assisted fuzzing of microservices is an emergent topic. 
Fig. ~\ref{fig:slr_overview} presents the distribution of the selected studies, showing both the yearly publication trend and the proportion of document types in the corpus. Venues include top-tier software engineering conferences (e.g., ICSE, FSE), specialized journals (e.g., IEEE TIFS, IEEE TSC), domain-specific venues and arXiv pre-prints, capturing the fast-evolving nature of the field. In terms of system focus, the corpus is dominated by REST-based APIs with limited representation of GraphQL, CGI/IoT services, and hybrid architectures, suggesting a narrow empirical base. 
Notably, only a small number of studies report industrial-scale cases, e.g., ByteDance and Volkswagen, while the majority rely on benchmark or single-service setups, highlighting a gap between current research and real-world multi-service environment practices.
The full list of selected primary studies is shown in the Appendix.

\subsection{RQ1: How are LLMs applied to support fuzzing in MSS?}
\label{sec:rq1}

We analyze RQ1 along four dimensions: \emph{which LLMs}, \emph{which roles} they play in the pipeline, \emph{which fuzzing visibility} they assume, and \emph{how} they are integrated.

\subsubsection{LLMs in Use.}

As shown in Table~\ref{tab:llms}, GPT-family models are the most common choice, but open-source and smaller models are also used, including Llama, Mistral, Qwen, DeepSeek, CodeBERT, and fine-tuned or quantised Llama variants. This diversity suggests that the field has not converged on a single canonical model; instead, reproducibility, cost, and local deployment increasingly influence model selection.

\begin{table}[t]
\centering
\caption{Models used in each primary study.}
\label{tab:llms}
\renewcommand{\arraystretch}{1.15}
\footnotesize

\hyphenpenalty=10000
\exhyphenpenalty=10000

\begin{tabular}{>{\raggedright\arraybackslash}p{2.5cm}
                >{\raggedright\arraybackslash}p{3.2cm}
                >{\raggedright\arraybackslash}p{6.3cm}}
\toprule
\textbf{Category} & \textbf{Studies} & \textbf{Examples} \\
\midrule

GPT family 
& \ref{SP1}, \ref{SP2}, \ref{SP3}, \ref{SP4}, \ref{SP6}, \ref{SP11}, \ref{SP12}, \ref{SP13}, \ref{SP14}, \ref{SP15}, \ref{SP16}, \ref{SP17}, \ref{SP19}

& GPT-3.5 / 3.5-turbo; GPT-4 family (4, 4o, 4o-mini, 4.1-mini); GPT-5 Mini; GPT-5.2-Codex; o1 / o1-mini; Azure OpenAI \\

\midrule

Other proprietary 
& \ref{SP2}, \ref{SP4}, \ref{SP15}
& Claude (3.5/3.7 Sonnet; Opus 4.6); Gemini (1.5, 2.0, 2.5) \\

\midrule

Open-source models 
& \ref{SP2}, \ref{SP6}, \ref{SP8}, \ref{SP19}, \ref{SP20}
& Llama 3 / 3.1 (8B, 70B); DeepSeek V3.1 \& R1; Mistral-Instruct 7B; Phi-3 Mini Instruct; WizardLM v1.2 13B \\

\midrule

Fine-tuned / Quantised 
& \ref{SP7}, \ref{SP18}
& CodeBERT; LlamaREST-IPD/EX \\

\midrule

Other / unspecified 
& \ref{SP5}, \ref{SP9}, \ref{SP10}
& StarCoder / StarCoderChat; Doubao-pro-128k; Not specified \\

\bottomrule
\end{tabular}
\end{table}

\subsubsection{LLM Role in the Fuzzing Pipeline.}
As shown in Table~\ref{tab:role-matrix}, LLMs are predominantly used as input generators, especially in black-box settings (16 studies), compared to a much smaller number in grey-box (1 study) and white-box settings (1 study). In contrast, the use of LLMs as mutation operators is limited, with only a few studies exploring this role in grey-box (1 study) and white-box (1 study) settings. Similarly, LLM-based oracles appear in both black-box (6 studies) and grey-box settings (2 studies), but remain relatively underexplored overall. Agent-based approaches employ LLMs as orchestrators in five studies in black-box settings, with little presence in other settings. Overall, the results show a clear concentration of LLM usage in input generation for black-box fuzzing, while their adoption in other roles are explored much less, and grey-box and white-box settings remain relatively underrepresented.


\begin{table}[!ht]
\centering
\caption{LLM roles $\times$ fuzzing visibility (one paper may appear in multiple cells).}
\label{tab:role-matrix}
\renewcommand{\arraystretch}{1.15}
\footnotesize

\hyphenpenalty=10000
\exhyphenpenalty=10000

\begin{tabular}{@{}
>{\raggedright\arraybackslash}p{2.6cm}
>{\raggedright\arraybackslash}p{5.6cm}
>{\raggedright\arraybackslash}p{1.8cm}
>{\raggedright\arraybackslash}p{1.8cm}
@{}}
\toprule
\textbf{LLM role} & \textbf{Black-box} & \textbf{Grey-box} & \textbf{White-box} \\
\midrule

Specification Generator 
& \ref{SP11}, \ref{SP16}, \ref{SP17}
& --
& -- \\

Input Generator 
& \ref{SP1}, \ref{SP2}, \ref{SP4}, \ref{SP5}, \ref{SP6}, \ref{SP8}, \ref{SP10}, \ref{SP11}, \ref{SP12}, \ref{SP13}, \ref{SP15}, \ref{SP16}, \ref{SP17}, \ref{SP18}, \ref{SP19}, \ref{SP20}
& \ref{SP9}
& \ref{SP14} \\

Mutation Operator 
& \ref{SP10}, \ref{SP17}, \ref{SP20}
& \ref{SP7}
& \ref{SP14} \\

Oracle / Analyzer 
& \ref{SP1}, \ref{SP2}, \ref{SP4}, \ref{SP10}, \ref{SP18}, \ref{SP19}
& \ref{SP7}, \ref{SP9}
& \ref{SP14} \\

Orchestrator 
& \ref{SP2}, \ref{SP6}, \ref{SP8}, \ref{SP16}, \ref{SP19}
& --
& -- \\

\bottomrule
\end{tabular}
\end{table}

\begin{table}[!ht]
  \centering
  \caption{Integration approaches (grouped by dimension).}
  \label{tab:integration-approach}
  \renewcommand{\arraystretch}{1.15}
  \footnotesize
  \begin{tabular}{@{}p{0.55\linewidth}p{0.44\linewidth}@{}}
    \toprule
    \textbf{Integration approach} & \textbf{Studies} \\
    \midrule

    \multicolumn{2}{@{}l}{\textit{Prompting strategy}} \\
    \quad One-shot generation
      & \ref{SP5}, \ref{SP10}, \ref{SP15} \\
    \quad Execution feedback loop
      & \ref{SP4}, \ref{SP11}, \ref{SP16}, \ref{SP17}, \ref{SP20} \\

    \addlinespace
    \multicolumn{2}{@{}l}{\textit{Architecture}} \\
    \quad Multi-agent
      & \ref{SP1}, \ref{SP6}, \ref{SP8}, \ref{SP16}, \ref{SP19} \\

    \addlinespace
    \multicolumn{2}{@{}l}{\textit{Knowledge augmentation}} \\
    \quad Retrieval-augmented generation (RAG)
      & \ref{SP2}, \ref{SP8} \\
    \quad Program analysis
      & \ref{SP7}, \ref{SP9}, \ref{SP14}, \ref{SP20} \\
    \quad Traffic / log / specification mining
      & \ref{SP9}, \ref{SP15}, \ref{SP17} \\

    \addlinespace
    \multicolumn{2}{@{}l}{\textit{Learning}} \\
    \quad Reinforcement learning (incl.\ multi-agent)
      & \ref{SP2}, \ref{SP12}, \ref{SP13}, \ref{SP18} \\
    \quad Fine-tuning / quantization
      & \ref{SP7}, \ref{SP18} \\

    \bottomrule
  \end{tabular}
\end{table}

\subsubsection{Integration Approach.} Table~\ref{tab:integration-approach} summarizes how LLMs are integrated into fuzzing pipelines across different dimensions. Prompt-based integration is the most common pattern, with iterative approaches using feedback being more prevalent than one-shot generation. Several recent studies adopt multi-agent architectures. For knowledge augmentation, program analysis (4 studies) and system data such as logs or traffic (3 studies) are more commonly used than retrieval-based approaches (2 studies). In contrast, learning-based integration is less common overall, with only limited use of reinforcement learning or model adaptation techniques.



\subsection{RQ2: How are LLM-based microservice fuzzers evaluated?}
\label{sec:rq2}

RQ2 covers target systems, metrics, baselines, and reported effectiveness, consolidating the per-study evaluation profile.

\subsubsection{Target Systems.}

Table~\ref{tab:targets} groups the evaluation targets into three categories: open benchmark services, deliberately vulnerable APIs, and industrial or real-world systems. However, most studies focus on evaluating individual REST services rather than cooperating microservice clusters, leaving cross-service workflows underexplored.

\begin{table}[t]
\centering
\caption{Target systems used in the primary studies.}
\label{tab:targets}
\renewcommand{\arraystretch}{1.15}
\footnotesize

\hyphenpenalty=10000
\exhyphenpenalty=10000

\begin{tabular}{@{}
>{\raggedright\arraybackslash}p{3cm}
>{\raggedright\arraybackslash}p{3.2cm}
>{\raggedright\arraybackslash}p{5.6cm}
@{}}
\toprule
\textbf{Category} & \textbf{Studies} & \textbf{Target systems} \\
\midrule

Open benchmark services
& \ref{SP11}, \ref{SP12}, \ref{SP13}, \ref{SP16}, \ref{SP18}, \ref{SP19}, \ref{SP20}
& FDIC, OhSome, Spotify, LanguageTool, RestCountries, Genome-Nexus, GitLab, PetStore \\

\midrule

Deliberately vulnerable APIs
& \ref{SP8}, \ref{SP11}
& VAmPI, CrAPI, OWASP Juice Shop \\

\midrule

Industrial / real-deployment systems
& \ref{SP5}, \ref{SP6}, \ref{SP7}, \ref{SP9}, \ref{SP11}, \ref{SP15}
& ByteDance microservices; Volkswagen Group~IT REST APIs; in-vehicle SPAPI; Light-OAuth2 microservice cluster; 20 Android backends; Azure/AWS/GCP cloud APIs \\

\bottomrule
\end{tabular}
\end{table}

\subsubsection{Metrics.}
Table~\ref{tab:metrics} shows that evaluation metrics cluster into four families: coverage, effectiveness, input quality, and cost. Coverage and effectiveness metrics are the only families reported in a comparable form across most studies therefore provide the common ground for cross-study comparison. Input quality metrics are concentrated in studies that focus on parameter or sequence generation, while cost-related metrics are reported by fewer than half of the studies. Overall, the breadth illustrates that the community has not yet converged on a benchmark suite, and the diversity of metrics remains a key obstacle to analysis.

\begin{table}[!ht]
\centering
\caption{Evaluation metrics reported in the primary studies.}
\label{tab:metrics}
\renewcommand{\arraystretch}{1.15}
\footnotesize

\hyphenpenalty=10000
\exhyphenpenalty=10000

\begin{tabular}{@{}
>{\raggedright\arraybackslash}p{2.6cm}
>{\raggedright\arraybackslash}p{3.2cm}
>{\raggedright\arraybackslash}p{6cm}
@{}}
\toprule
\textbf{Family} & \textbf{Studies} & \textbf{Metrics} \\
\midrule

Coverage-based
& \ref{SP1}, \ref{SP2}, \ref{SP3}, \ref{SP5}, \ref{SP11}, \ref{SP12}, \ref{SP13}, \ref{SP14}, \ref{SP15}, \ref{SP16}, \ref{SP18}
& Line, branch, and method coverage; API/operation coverage; schema coverage; basic-block coverage; log-template coverage \\

\midrule

Effectiveness
& \ref{SP2}, \ref{SP3}, \ref{SP6}, \ref{SP7}, \ref{SP9}, \ref{SP10}, \ref{SP11}, \ref{SP12}, \ref{SP13}, \ref{SP16}, \ref{SP17}, \ref{SP18}
& Number of confirmed bugs or vulnerabilities; unique 5xx server errors; 500-error counts as proxy oracle; CVEs assigned; pass rate and fault detection; root-cause categorisation \\

\midrule

Input quality
& \ref{SP5}, \ref{SP8}, \ref{SP10}, \ref{SP11}, \ref{SP15}, \ref{SP17}, \ref{SP18}
& Valid-input/pass rate; valid-sequence increase; value-generation precision; inter-parameter dependency (IPD) detection accuracy; false-positive parameter rate; syntactic correctness of generated tests \\

\midrule

Cost / engineering
& \ref{SP1}, \ref{SP4}, \ref{SP6}, \ref{SP8}, \ref{SP9}, \ref{SP14}, \ref{SP15}, \ref{SP16}, \ref{SP17}
& Token usage; processing time; GPU/CPU speedup; energy consumption; deployment effort; generation and execution time \\

\bottomrule
\end{tabular}
\end{table}

\subsubsection{Baselines.}
Table~\ref{tab:baselines} shows that traditional fuzzers are the dominant baseline. A small group of studies compares only against other LLMs which weakens the strength of their effectiveness claims. Three studies compare against \emph{human-written} test suites or expert ground truth, an important sanity check that is otherwise rare. In addition, replication packages are released by 11 out of 20 studies (55\%), which facilitates reproducibility and enables further empirical validation.

\begin{table}[!ht]
\centering
\caption{Baselines and replication artefacts in the primary studies.}
\label{tab:baselines}
\renewcommand{\arraystretch}{1.15}
\footnotesize

\hyphenpenalty=10000
\exhyphenpenalty=10000

\begin{tabular}{@{}
>{\raggedright\arraybackslash}p{3.2cm}
>{\raggedright\arraybackslash}p{2.8cm}
>{\raggedright\arraybackslash}p{5.6cm}
@{}}
\toprule
\textbf{Baseline category} & \textbf{Studies} & \textbf{Details} \\
\midrule

Traditional fuzzers
& Most studies
& RESTler (9); EvoMaster (8); Morest (5); ARAT-RL (4); foREST, RESTTESTGEN, RESTCT (1) \\

\midrule

LLM-only comparisons
& \ref{SP4}, \ref{SP8}, \ref{SP19}
& Compared only against other LLMs, limiting the strength of effectiveness claims \\

\midrule

Human-written / expert ground truth
& \ref{SP5}, \ref{SP6}, \ref{SP15}
& Compared against human-written test suites or expert ground truth, providing an important but rare sanity check \\

\midrule

Replication packages
& 11/20 (55\%)
& Generally good release rate supporting reproducibility; studies \ref{SP19} and \ref{SP20} do not provide artefacts \\

\bottomrule
\end{tabular}
\end{table}


\subsection{RQ3: Open challenges and future research directions.}
\label{sec:rq3}

As shown in Table~\ref{tab:challenges}, the open challenges of LLM-based microservice fuzzing reflect a shift from feasibility-oriented prototypes to deployable testing systems. Methodologically, current approaches still struggle with guarded branches, hallucinated or invalid tests, and complex inter-parameter or OpenAPI dependencies, suggesting that prompt-based generation alone is insufficient. Evaluation remains limited by the absence of unified microservice benchmarks and by the frequent reliance on weak oracles such as 5xx responses, which cannot fully capture functional or security failures. From an engineering perspective, token cost, latency, compute overhead, non-determinism, and IP/compliance concerns directly hinder CI/CD integration. At the theoretical level, existing systems provide limited guarantees that LLM-generated tests satisfy API grammars, protocol constraints, or distributed authorisation logic.
\begin{table}[!ht]
\centering
\caption{Open challenges and corresponding future-work directions.}
\label{tab:challenges}
\renewcommand{\arraystretch}{1.2}
\footnotesize

\hyphenpenalty=10000
\exhyphenpenalty=10000

\begin{tabular}{@{}
>{\raggedright\arraybackslash}p{2cm}
>{\raggedright\arraybackslash}p{5cm}
>{\raggedright\arraybackslash}p{5cm}
@{}}
\toprule
\textbf{Dimension} & \textbf{Open challenges} & \textbf{Future directions} \\
\midrule

Methodology 
& Limited coverage on guarded branches; hallucinated or invalid tests; weak handling of inter-parameter dependencies and incomplete specifications 
(\ref{SP1}, \ref{SP5}, \ref{SP8}, \ref{SP10}, \ref{SP11}, \ref{SP12}, \ref{SP14}, \ref{SP16}, \ref{SP17}, \ref{SP18}, \ref{SP19}, \ref{SP20})
& Hybrid LLM + static/symbolic analysis; self-reflective feedback loops with LLM-as-judge oracles; IPD-aware fine-tuning and specification mining 
(\ref{SP1}, \ref{SP2}, \ref{SP11}, \ref{SP14}, \ref{SP15}, \ref{SP16}, \ref{SP17}, \ref{SP18}, \ref{SP19}, \ref{SP20}) \\

\midrule

Evaluation 
& Lack of unified microservice benchmarks; reliance on weak 5xx-only oracles 
(\ref{SP8}, \ref{SP15}, \ref{SP16}, \ref{SP17})
& Standardised authorisation and log-coverage benchmarks; functional and response-difference oracles 
(\ref{SP8}, \ref{SP15}, \ref{SP16}, \ref{SP17}) \\

\midrule

Engineering 
& High compute, token, and latency cost; limited CI/CD integration; non-determinism and IP/compliance concerns 
(\ref{SP1}, \ref{SP2}, \ref{SP5}, \ref{SP8}, \ref{SP9}, \ref{SP14}, \ref{SP15}, \ref{SP17}, \ref{SP18}, \ref{SP19})
& Smaller, quantised, or domain-specific LLMs; on-prem deployment; deterministic decoding; CI/CD-friendly tooling 
(\ref{SP1}, \ref{SP2}, \ref{SP5}, \ref{SP9}, \ref{SP15}, \ref{SP18}, \ref{SP20}) \\

\midrule

Theory 
& Lack of formal correctness guarantees; difficulty modelling distributed authorisation logic 
(\ref{SP5}, \ref{SP7}, \ref{SP8}, \ref{SP10})
& Constrained, grammar-guided generation; verified post-conditions; cross-API dependency reasoning 
(\ref{SP2}, \ref{SP7}, \ref{SP8}, \ref{SP10}) \\

\bottomrule
\end{tabular}
\end{table}

Overall, the result suggests that future research should move beyond simply adding LLMs to fuzzing pipelines. Progress will depend on three directions: microservice-aware designs that combine LLMs with static or symbolic analysis, dependency graphs, runtime feedback, and cross-API reasoning; cost-aware deployment strategies based on smaller, quantised, or domain-specific models with deterministic decoding; and correctness-aware generation methods such as grammar-guided decoding, verified post-conditions, and stronger functional or security oracles. In this sense, the next stage of LLM-assisted microservice fuzzing should focus not only on generating more realistic inputs, but also on making the resulting systems scalable, reproducible, and suitable for industrial deployment.

\section{Discussion}
\label{sec:disc}

The synthesis in Section \ref{sec:results} reveals several key observations about the current state of LLM-assisted fuzz testing in MSS and its implications for research and practice.

\textbf{Architectural convergence.} The field is converging on black-box, feedback-driven, and agentic fuzzing pipelines that combine LLM-based input generation, response/log analysis, and lightweight orchestration. This shared template helps cumulative progress, but pure black-box pipelines may soon face diminishing returns without richer external signals such as source code, dependency graphs, runtime traces, or service-level observability.

\textbf{Microservice realism gap.} Most evaluations still target isolated REST services rather than cooperating microservice clusters. Cross-service workflows, distributed authorisation, asynchronous queues, service-mesh policies, and sidecar interactions are rarely tested. Therefore, reported gains in validity, coverage, or bug discovery may not fully reflect performance in production microservice environments.

\textbf{Cost and deployability.} Token cost, latency, energy consumption, non-determinism, and compliance concerns remain major barriers to CI/CD adoption. This makes smaller, quantised, fine-tuned, or domain-specific models increasingly attractive, especially when they can run on-premise. Model downsizing is thus a deployment requirement, not merely an optimisation.

\textbf{Implications.} Practitioners should combine LLM fuzzing with program analysis or dependency information when available, and prefer locally deployable models when cost or compliance is critical. Researchers should prioritise microservice-cluster benchmarks, cost-aware evaluation, and correctness-aware generation using grammars, specifications, or verified post-conditions.

\section{Threats to Validity}
\label{sec:threats}

Here, we discuss threats to validity following the categorization proposed by Ampatzoglou et al. that fits software engineering secondary studies, covering study selection, data and research validity \cite{ampatzoglou2019identifying}.

\textbf{Study-selection validity.} The completeness of the selected corpus may be affected by the search strategy and source coverage. 
Although we queried major digital libraries, e.g., IEEE Xplore, ACM DL, Scopus, WoS, and complemented them with snowballing, relevant studies may have still been missed. We include pre-prints from ArXiv to capture the fast-evolving nature of this research area, their non-peer-reviewed status may introduce variability in study quality and maturity. 
The inclusion and exclusion criteria were designed to align closely with the research scope, though borderline cases required subjective judgment. 
To mitigate this, a dual-reviewer screening process with conflict resolution was applied, but residual selection bias may remain.

\textbf{Data validity.} Threats arise from potential inaccuracies or inconsistencies in data extraction and classification. Hence, categorical fields (LLM role, fuzzing type, integration approach) were extracted using a controlled vocabulary. Nonetheless, some studies report hybrid or implicit configurations, which require the interpretation of the reviewers. 
Furthermore, heterogeneity in reporting, e.g., metrics, model details, or experimental setups, limits direct comparability across studies and may introduce bias in synthesis. To reduce errors, data extraction was performed by one researcher and cross-checked by another, although minor inconsistencies may persist.

\textbf{Research validity.} The synthesis and interpretation of results are subject to researcher bias, particularly in taxonomy design and qualitative aggregation. The synthesis was performed primarily by a single reviewer with cross-checking despite the fact that three reviewers were involved in extraction. Although the protocol-driven methodology aligned with the SLR guidelines, the categorization of LLM roles, integration patterns, and evaluation practices reflects informed but subjective decisions. 
Furthermore, the lack of standardized benchmarks and metrics across primary studies constrains the strength of generalizable conclusions. To mitigate these threats, we focused on consistent patterns across multiple studies and avoided overgeneralization of isolated results. 
The corpus skews towards REST-based systems with minimal representation of GraphQL, RPC, and other API paradigms, which constrains the applicability of conclusions to broader microservice ecosystems. 
Industrial evidence is concentrated in a small number of large-scale case studies, which may not generalize to organizations with different scales, technology stacks, or regulatory contexts. To mitigate these threats, we focus on recurring patterns across multiple studies and avoid overgeneralizing isolated results. 

\section{Conclusion}
\label{sec:conclusion}

We presented a systematic literature review of LLM-assisted fuzz testing for microservice systems, synthesizing 20 primary studies published between 2024 and 2026. The results show a clear convergence where LLMs are mainly used as semantic input generators in black-box fuzzing, with increasing adoption of closed-loop, 
multi-agent and RAG-augmented architectures that incorporate oracle and orchestration roles.
Meanwhile, LLMs improve valid input generation and modestly enhance coverage and vulnerability detection, but evaluation remains inconsistent, with limited cost reporting and a bias toward single-service experiments over realistic microservice systems. 
Notably, the key challenges persist across literature including the cost–effectiveness trade-off, lack of standardized benchmarks, weak oracles, and no formal guarantees. The field is converging on three directions: (1) smaller or fine-tuned models for efficiency, (2) hybrid LLM–program analysis approaches, and (3) standardized, cost-aware multi-service benchmarks. 
Overall, although LLMs significantly enhance fuzzing, integration into scalable, reproducible, and industry-ready pipelines requires more realistic microservice-level evaluation frameworks. 

\section{Disclosure of Interests}
The authors have no competing interests to declare that are relevant to the content of this article.

\section{Acknowledgements}
This work is supported by FAST, the Finnish Software Engineering Doctoral Research Network, funded by the Ministry of Education and Culture, Finland. This work is also supported by the Research Council of Finland (grant id: 359861, the MuFAno project). This work is also funded by the EuroHPC Joint Undertaking and its members including top-up funding by the Ministry of Education and Culture.

\appendix
\section*{Appendix - The Selected Papers (SP$_s$)}
\label{appendix}

{\small
\begin{enumerate}[labelindent=0pt,leftmargin=*,label={[SP\arabic*]}]

\item \label{SP1}
Nooyens, R., Bardakci, T., Beyazıt, M., \& Demeyer, S. (2025, September). Test amplification for rest apis via single and multi-agent llm systems. In IFIP International Conference on Testing Software and Systems (pp. 161-177). Cham: Springer Nature Switzerland.

\item \label{SP2}
Liu, S., Marefat, S., Tsai, O., Chen, Y., Deng, Z., Wang, J., \& Tayebi, M. A. (2025). PrediQL: Automated Testing of GraphQL APIs with LLMs. arXiv preprint arXiv:2510.10407.

\item \label{SP3}
Shi, C., Yu, J., Zhao, Z., Chen, J., \& Zhang, F. (2025). CGIFuzz: Enabling Gray-Box Fuzzing for Web CGI of IoT Devices. IEEE Transactions on Information Forensics and Security.

\item \label{SP4}
Altin, M., Mutlu, B., Kilinc, D., \& Cakir, A. (2025). Automated Testing for Service-Oriented Architecture: Leveraging Large Language Models for Enhanced Service Composition. IEEE Access.

\item \label{SP5}
Poth, A., Rrjolli, O., \& Arcuri, A. (2025). Technology adoption performance evaluation applied to testing industrial REST APIs. Automated Software Engineering, 32(1), 5.

\item \label{SP6}
Wang, S., Yu, Y., Feldt, R., \& Parthasarathy, D. (2025). Automating a complete software test process using llms: An automotive case study. arXiv preprint arXiv:2502.04008.

\item \label{SP7}
Yijun, L., Shuhao, L., Zufeng, W., Zhizhong, W., Jianwen, H., \& Lei, Z. (2025, December). SemFuzz: A Semantic-Guided Fuzzing Framework for Detecting Unauthorized API Access in Mobile Applications. In 2025 22nd International Computer Conference on Wavelet Active Media Technology and Information Processing (ICCWAMTIP) (pp. 1-5). IEEE.

\item \label{SP8}
Pasca, E. M., Delinschi, D., Erdei, R., \& Matei, O. (2025). LLM-Driven, Self-Improving Framework for Security Test Automation: Leveraging Karate DSL for Augmented API Resilience. IEEE Access.

\item \label{SP9}
Wang, J., Chen, S., Liu, Y., Deng, Y., Zhang, L., Fu, Y., \& Liu, B. (2025, June). TestGPT-Server: Automatically Testing Microservices with Large Language Models at ByteDance. In Proceedings of the 33rd ACM International Conference on the Foundations of Software Engineering (pp. 192-203).

\item \label{SP10}
Chen, J., Chen, Y., Pan, Z., Chen, Y., Li, Y., Li, Y., ... \& Shen, Y. (2024). Dyner: Optimized test case generation for representational state transfer (rest) ful application programming interface (api) fuzzers guided by dynamic error responses. Electronics, 13(17), 3476.

\item \label{SP11}
Zheng, T., Shao, J., Dai, J., Jiang, S., Chen, X., \& Shen, C. (2024). Restless: Enhancing state-of-the-art rest api fuzzing with llms in cloud service computing. IEEE Transactions on Services Computing, 17(6), 4225-4238.

\item \label{SP12}
Stennett, T., Kim, M., Sinha, S., \& Orso, A. (2025, April). Autoresttest: A tool for automated rest api testing using llms and marl. In 2025 IEEE/ACM 47th International Conference on Software Engineering: Companion Proceedings (ICSE-Companion) (pp. 21-24). IEEE.

\item \label{SP13}
Kim, M., Stennett, T., Sinha, S., \& Orso, A. (2025, April). A Multi-Agent Approach for REST API Testing with Semantic Graphs and LLM-Driven Inputs. In Proceedings of the IEEE/ACM 47th International Conference on Software Engineering (pp. 1409-1421).

\item \label{SP14}
Li J., Shen J., Su Y., \& Lyu M. R. (2026). MioHint: LLM-assisted Mutation for Whitebox API Testing. arXiv preprint arXiv:2504.05738 .

\item \label{SP15}
Reinikainen, N., Mäntylä, M., \& Wang, Y. (2026). Assessing REST API Test Generation Strategies with Log Coverage. arXiv preprint arXiv:2604.07073.

\item \label{SP16}
Kogler, L., Ehrhart, M., Dornauer, B., \& Enoiu, E. P. (2025). RESTifAI: LLM-Based Workflow for Reusable REST API Testing. arXiv preprint arXiv:2512.08706.

\item \label{SP17}
Decrop, A., Devroey, X., Papadakis, M., Schobbens, P. Y., \& Perrouin, G. (2024). You Can REST Now: Automated REST API Documentation and Testing via LLM-Assisted Request Mutations. arXiv preprint arXiv:2402.05102.

\item \label{SP18}
Kim, M., Sinha, S., \& Orso, A. (2025). Llamaresttest: Effective rest api testing with small language models. Proceedings of the ACM on Software Engineering, 2(FSE), 465-488.

\item \label{SP19}
Han, X., \& Zhu, H. (2025). MASTEST: A LLM-Based Multi-Agent System For RESTful API Tests. arXiv preprint arXiv:2511.18038.

\item \label{SP20}
Liu, C. H., Chen, S. L., \& Li, K. Y. (2026). REST API Fuzzing Using API Dependencies and Large Language Models. Engineering Proceedings, 120(1), 42.


\end{enumerate}
}

\bibliographystyle{splncs04}
\bibliography{bib}

@article{ampatzoglou2019identifying,
  title={Identifying, categorizing and mitigating threats to validity in software engineering secondary studies},
  author={Ampatzoglou, Apostolos and Bibi, Stamatia and Avgeriou, Paris and Verbeek, Marijn and Chatzigeorgiou, Alexander},
  journal={Information and software technology},
  volume={106},
  pages={201--230},
  year={2019},
  publisher={Elsevier}
}

@article{kitchenham2009systematic,
  title={Systematic literature reviews in software engineering--a systematic literature review},
  author={Kitchenham, Barbara and Brereton, O Pearl and Budgen, David and Turner, Mark and Bailey, John and Linkman, Stephen},
  journal={Information and software technology},
  volume={51},
  number={1},
  pages={7--15},
  year={2009},
  publisher={Elsevier}
}

@article{wohlin2014guidelines,
  title={Guidelines for snowballing in systematic literature studies and a replication in software engineering},
  author={Wohlin, Claes},
  journal={Proceedings of the 18th international conference on evaluation and assessment in software engineering},
  pages={1--10},
  year={2014}
}

@article{golmohammadi2023testing,
  title={Testing RESTful APIs: A survey},
  author={Golmohammadi, Amid and Zhang, Man and Arcuri, Andrea},
  journal={ACM Transactions on Software Engineering and Methodology},
  volume={33},
  number={1},
  pages={1--41},
  year={2023}
}

@article{manes2019art,
  title={The art, science, and engineering of fuzzing: A survey},
  author={Manes, Valentin Jean Marie and Han, HyungSeok and Han, Choongwoo and Cha, Sang Kil and Egele, Manuel and Schwartz, Edward J and Woo, Maverick},
  journal={IEEE Transactions on Software Engineering},
  volume={47},
  number={11},
  pages={2312--2331},
  year={2019}
}

@article{kaniewski2025systematic,
  title={A systematic literature review on detecting software vulnerabilities with large language models},
  author={Kaniewski, Sabrina and Schmidt, Fabian and Enzweiler, Markus and Menth, Michael and Heer, Tobias},
  journal={arXiv preprint arXiv:2507.22659},
  year={2025}
}

@article{he2024large,
  title={Large language models for blockchain security: A systematic literature review},
  author={He, Zheyuan and Li, Zihao and Yang, Sen and Ye, He and Qiao, Ao and Zhang, Xiaosong and Luo, Xiapu and Chen, Ting},
  journal={arXiv preprint arXiv:2403.14280},
  year={2024}
}

@article{ghani2019microservice,
  author  = {Ghani, Israr and Wan-Kadir, Wan M. N. and Mustafa, Ahmad and Imran Babir, Muhammad},
  title   = {Microservice Testing Approaches: A Systematic Literature Review},
  journal = {International Journal of Integrated Engineering},
  volume  = {11},
  number  = {8},
  pages   = {65--80},
  year    = {2019},
  doi     = {10.30880/ijie.2019.11.08.008}
}

@article{xu2025largelanguagemodelscyber,
author = {Xu, Hanxiang and Wang, Shenao and Li, Ningke and Wang, Kailong and Zhao, Yanjie and Chen, Kai and Yu, Ting and Liu, Yang and Wang, Haoyu},
title = {Large Language Models for Cyber Security: A Systematic Literature Review},
year = {2025},
publisher = {Association for Computing Machinery},
address = {New York, NY, USA},
issn = {1049-331X},
url = {https://doi.org/10.1145/3769676},
doi = {10.1145/3769676},
note = {Just Accepted},
journal = {ACM Trans. Softw. Eng. Methodol.},
month = sep
}

@article{PONCE2025107870,
title = {Microservices testing: A systematic literature review},
journal = {Information and Software Technology},
volume = {188},
pages = {107870},
year = {2025},
issn = {0950-5849},
doi = {https://doi.org/10.1016/j.infsof.2025.107870},
url = {https://www.sciencedirect.com/science/article/pii/S0950584925002095},
author = {Francisco Ponce and Roberto Verdecchia and Breno Miranda and Jacopo Soldani}
}

@INPROCEEDINGS{Waseem,
  author={Waseem, Muhammad and Liang, Peng and Márquez, Gastón and Salle, Amleto Di},
  booktitle={2020 27th Asia-Pacific Software Engineering Conference (APSEC)}, 
  title={Testing Microservices Architecture-Based Applications: A Systematic Mapping Study}, 
  year={2020},
  volume={},
  number={},
  pages={119-128},
  doi={10.1109/APSEC51365.2020.00020}}

@inproceedings{S9_Wang2025bytedance,
  title={{TestGPT-Server}: Automatically Testing Microservices with Large Language Models at {ByteDance}},
  author={Wang, J. and Chen, S. and Liu, Y. and Deng, Y. and Zhang, L. and Fu, Y. and Liu, B.},
  booktitle={Proc. ACM SIGSOFT Symposium on the Foundations of Software Engineering (FSE)},
  year={2025},
  doi={10.1145/3696630.3728545}
}

@misc{S15_LogCoverage2026,
  title={Assessing {REST} {API} Test Generation Strategies with Log Coverage},
  author={Reinikainen, Nana and M{\"a}ntyl{\"a}, Mika and Wang, Yuqing},
  year={2026},
  howpublished={arXiv:2604.07073}
}

@INPROCEEDINGS{Huang2025,
  author={Huang, Linghan and Zhao, Peizhou and Ma, Lei and Chen, Huaming},
  booktitle={2025 IEEE International Conference on Software Services Engineering (SSE)}, 
  title={On the Challenges of Fuzzing Techniques via Large Language Models}, 
  year={2025},
  volume={},
  number={},
  pages={162-171},
  doi={10.1109/SSE67621.2025.00028}}

@article{dragoni2017microservices,
  title={Microservices: yesterday, today, and tomorrow},
  author={Dragoni, Nicola and Giallorenzo, Saverio and Lafuente, Alberto Lluch and Mazzara, Manuel and Montesi, Fabrizio and Mustafin, Ruslan and Safina, Larisa},
  journal={Present and ulterior software engineering},
  pages={195--216},
  year={2017},
  publisher={Springer}
}

@article{wang2021promises,
  title={Promises and challenges of microservices: an exploratory study},
  author={Wang, Yingying and Kadiyala, Harshavardhan and Rubin, Julia},
  journal={Empirical Software Engineering},
  volume={26},
  number={4},
  pages={63},
  year={2021},
  publisher={Springer}
}

@article{miao2025systematic,
  title={Systematic Mapping Study of Test Generation for Microservices: Approaches, Challenges, and Impact on System Quality},
  author={Miao, Tingshuo and Shaafi, Asif Imtiaz and Song, Eunjee},
  journal={Electronics},
  volume={14},
  number={7},
  pages={1397},
  year={2025},
  publisher={MDPI}
}

@article{zhang2023open,
  title={Open problems in fuzzing restful apis: A comparison of tools},
  author={Zhang, Man and Arcuri, Andrea},
  journal={ACM Transactions on Software Engineering and Methodology},
  volume={32},
  number={6},
  pages={1--45},
  year={2023},
  publisher={ACM New York, NY}
}

@inproceedings{deng2023large,
  title={Large language models are zero-shot fuzzers: Fuzzing deep-learning libraries via large language models},
  author={Deng, Yinlin and Xia, Chunqiu Steven and Peng, Haoran and Yang, Chenyuan and Zhang, Lingming},
  booktitle={Proceedings of the 32nd ACM SIGSOFT international symposium on software testing and analysis},
  pages={423--435},
  year={2023}
}
\end{document}